%% file: RCdataPP.tex
\documentclass[preprint]{aastex}
\usepackage{emulateapj5}

\begin{document}

\newcommand{\kms}{\ensuremath{\,\mbox{km}\,\mbox{s}^{-1}}}
\newcommand{\etal}{et al.}
\newcommand{\Msun}{\ensuremath{M_{\odot}}}
\newcommand{\Lsun}{\ensuremath{L_{\odot}}}
\newcommand{\HI}{H{\sc i}}
\newcommand{\MLs}{\ensuremath{\Upsilon_{\star}}}
\newcommand{\Ha}{H$\alpha$}
\newcommand{\Rth}{\ensuremath{R_{34}}}
\newcommand{\Vf}{\ensuremath{V_{flat}}}
\newcommand{\Vsys}{\ensuremath{V_{hel}}}

\shortauthors{McGaugh, Rubin, \& de Blok}
\shorttitle{LSB rotation curves}

\title{High-resolution rotation curves of low surface brightness galaxies:
Data}

\author{Stacy S.\ McGaugh\altaffilmark{1}}
\affil{Department of Astronomy, University of Maryland}
\affil{College Park, MD 20742-2421, USA}
\email{ssm@astro.umd.edu}

\author{Vera C.\ Rubin\altaffilmark{1}}
\affil{Department of Terrestrial Magnetism}
\affil{Carnegie Institution of Washington }
\affil{5241 Broad Branch Rd., N. W.}
\affil{Washington, D. C. 20015, USA}
\email{rubin@dtm.ciw.edu}

\and

\author{W.J.G.\ de Blok\altaffilmark{2}}
\affil{Australia Telescope National Facility}
\affil{PO Box 76, Epping NSW 1710, Australia}
\email{edeblok@atnf.csiro.au}

\altaffiltext{1}{Based on observations using the 4 m telescope,
 Kitt Peak National Observatory, National Optical
     Astronomy Observatories, which is operated by the Association of
     Universities for Research in Astronomy, Inc. (AURA) under cooperative
     agreement with the National Science Foundation, and the du Pont
telescope of the Las Campanas Observatory, Carnegie Institution of
Washington.}

\altaffiltext{2}{Bolton Fellow}

\begin{abstract}

We present long slit H$\alpha$ observations of 50 low surface brightness
galaxies.  Of these, 36 are of sufficient quality to form rotation curves.
These data provide a large
increase in the number of low surface brightness galaxies for which
accurate rotation curves are available.  They also represent an order
of magnitude improvement in spatial resolution over previous 21 cm
studies ($1''$ to $2''$ instead of $13''$ to $45''$).  The improved
resolution and accuracy of the data extend and strengthen the scientific
conclusions previously inferred from 21 cm data.

\end{abstract}


\section{Introduction}

The rotation curves of spiral galaxies contain important
information about the gravitational potential produced by their combined
mass components. They played a fundamental role in
establishing the need for dark matter (Sofue \& Rubin 2001).  The rotation
curves of low surface brightness (LSB) galaxies are of particular
interest because they offer evidence that LSB galaxies are dark matter 
dominated (de Blok \& McGaugh 1996, 1997; Pickering \etal\ 1997, 1999;
Blais-Ouellette, Amram, \& Carignan 2001).
Consequently, they have tremendous potential to constrain theories of
galaxy formation (McGaugh \& de Blok 1998a; van den Bosch \& Dalcanton
2000), to probe the nature of dark matter (Pfenniger, Combes, \&
Martinet 1994; Spergel \& Steinhardt 2000), and to test alternatives
to dark matter (McGaugh \& de Blok 1998b; de Blok \& McGaugh 1998).
 
 With central surface brightnesses $\mu_0^B > 22.7\;{\rm mag.}\;{\rm
arcsec}^{-2}$ (McGaugh 1996; Impey \& Bothun 1997; Bothun, Impey, \&
McGaugh 1997), the maximum stellar surface densities of LSB galaxies are
typically $< 100\ \Msun\;{\rm pc}^{-2}$ for plausible stellar
mass-to-light ratios (\MLs: Bell \& de Jong 2001). 
Yet their terminal rotation velocities (\Vf) are no
lower than those of higher surface brightness galaxies --- they adhere to
the same Tully-Fisher relation (Sprayberry \etal\ 1995; Zwaan
\etal\ 1995; Hoffman \etal\ 1996; see also Courteau \& Rix 1999). This 
lack of Tully-Fisher zero point dependence on surface brightness is 
widely interpreted to mean that dark matter dominates the observed
rotation at essentially all radii, so that the rotation curves
directly probe the invisible mass component.  Degeneracies
between stellar and dark mass which have plagued decompositions of the
rotation curves of high surface brightness galaxies (e.g., Kent 1987)
become much less severe in LSB galaxies.

In order to exploit this property of LSB galaxies to map out the dark
matter distribution, it is desirable
to obtain high spatial resolution data.  The \HI\ data presented by
de Blok, McGaugh, \& van der Hulst (1996) were adequate to demonstrate
many of the systematic properties which make LSB galaxy
rotation curves particularly
interesting.  However, the beam size of those observations ($\ge 13''$)
led to concern over the effects of beam smearing on the
derived shapes of the rotation curves (van den Bosch \etal\ 2000;
Swaters, Madore, \& Trewahla 2000). 
Though de Blok \& McGaugh (1997) had shown that beam smearing
was unlikely to be important in most (though not all)
cases, controversy persists because  these data are important in
testing the cuspy halos predicted in cosmological simulations with cold
dark matter (Navarro, Frenk, \& White 1997 [NFW]; Moore \etal\ 1999).  A
direct way to address these concerns is to improve the spatial resolution
of the observations.  To this end, we have obtained
optical rotation curves based on long slit observations of the H$\alpha$
emission line.  With seeing limited angular resolution of $\sim 1'' - 1.5''$,
corresponding to sub-kpc physical scales for the LSB galaxies in the
sample, these new data represent over an order of magnitude improvement in
spatial resolution.

For these new data, the error bars are often considerably
smaller than in previous 21 cm studies (van der Hulst \etal\ 1993;
de Blok \etal\ 1996).  This accuracy, as well as the improved
spatial resolution, are essential for improving constraints on detailed
mass models, which are presented in a companion paper
(de Blok, McGaugh, \& Rubin 2001; hereafter Paper II).
Our aim is to probe in detail the rotation curves, and hence the implied
mass distributions (de Blok, McGaugh, Bosma, \& Rubin 2001),
for a significant number of galaxies which appear to be dark matter
dominated.

This paper presents the H$\alpha$ rotation curve measurements.
The sample and the data are described in \S 2.  Section 3 provides
a direct comparison of these new data with available 21 cm data,
and with independent \Ha\ observations.  A summary of the
scientific impact of the new data is given in \S 4, and conclusions 
are in \S 5.

\section{The Data}

\subsection{Sample}

The principal targets for optical rotation curves are the LSB galaxies in
the \HI\ sample of de Blok \etal\ (1996) and van der Hulst \etal\ (1993),
with surface photometry and
colors as reported in those papers and by McGaugh \& Bothun (1994)
and de Blok, van der Hulst \& Bothun (1995).
These galaxies are blue, late type LSB galaxies
found in searches of deep photographic plates (Schombert
\etal\ 1992; Impey \etal\ 1996) and exemplify the regime of
low stellar mass surface density.  We also have observed a few of
the LSB dwarf galaxies of Schombert, Pildis, \& Eder (1997), which
are in may ways similar to the other LSB galaxies but are typically of
lower luminosity.  The sample does not include, as yet, red LSB galaxies
such as those of O'Neil \etal\ (1997) and O'Neil, Bothun, \& Schombert
(2000). In addition, we also observed
LSB galaxies from the UGC (Nilson 1973)
and ESO-LV (Lauberts \& Valentijn 1989)
catalogs as time permitted.  The ESO-LV catalog
includes surface photometry information, so selection was based
on central surface brightness:
$\mu_0^B > 23\;{\rm mag.}\;{\rm arcsec}^{-2}$.  
Of galaxies in the accessible range of right ascension
which met this criteria, we gave preference to ESO-LV
galaxies which have been included in photometric studies
(e.g., Matthews \& Gallagher 1997; Bell \etal\ 2000).

Galaxies selected from the UGC are estimated to be of low surface
brightness based on catalog diameters and magnitudes. 
Visual inspection of
their images in the digital sky survey confirmed their LSB
nature. For these, a  knotty OB
association morphology is a good predictor of ubiquitous H$\alpha$
emission.

\subsection{Observations}

The data presented here were obtained with the Las Campanas
du Pont 2.5 m telescope and the Kitt Peak 4 m telescope.
We list in Table 1 observations for galaxies for which
we have obtained velocity measurements, arranged by catalog:
LSBs, UGC, and ESO-LV.  
Successive columns list (1) the name of the galaxy, (2) the heliocentric
system velocity, (3) the disk inclination, (4) the
position angle of the observation, (5) the quality of the data,
(6) whether resolved \HI\ data exist, and (7) whether a mass model is
constructed in Paper II.  The quality flag denotes good (Q = 1),
fair (2), and poor (3) \Ha\ data.
It bears no relation to the quality of the \HI\
synthesis data in cases where it exists.  In general, the \Ha\ data are
better for defining the inner shape of the rotation curve, while the \HI\
data extend to larger radii.  

For mass modeling, the quality of the data must be good.  Every galaxy is
unique and not all data can be blindly used for any purpose  (see
discussions in de Blok \& McGaugh 1997 and McGaugh \& de Blok 1998a). 
Of the 26 galaxies (Table 1) which are modeled in Paper II, 25 have Q = 1
(good).

The \HI\ data are taken
from van der Hulst \etal\ (1993) and de Blok \etal\ (1996).
When available, the \HI\
data provide a good kinematic indication of the position angle.
This generally corresponds to the optical major axis of the galaxy, which
is used as the position angle when \HI\ data are not available.  Disk
inclinations are adopted from previous studies (van der Hulst \etal\ 1993;
de Blok \etal\ 1996) or are measured from optical axis ratios using
$\cos^2(i) = 1.042(b/a)^2 - 0.042$.  Photometric inclinations
for LSB galaxies are intrinsically uncertain because of their generally
ragged, late type morphologies and the low signal-to-noise in the outermost
isophotes (see de Blok \& McGaugh 1998 for a more extensive discussion
of inclination uncertainties.)  Generally, the tabulated values should be
good to better than $\pm 5^{\circ}$, though large excursions from that
can not be discounted in a few cases.  There is also some indication
that LSB galaxies tend to be thinner (Dalcanton \& Bernstein 2000;
Matthews \& van Driel 2000) than the edge-on axis ratio of $a/b = 5$
implied by the formula we use for $\cos^2(i)$.  This is a minor issue
compared to the intrinsic uncertainty in the observed axis ratios.
Fortunately, inclination errors only affect the absolute scale of
rotation velocities, and not the shape of the rotation curve.
Systemic velocities are derived from the \Ha\ data as described below.

\placetable{Table1}

\subsubsection{Las Campanas}

Observations of galaxies from the ESO-LV catalog were made 
with the Las Campanas 2.5 m telescope in November 1998.  
We used the modular spectrograph with a 600 line/mm grating blazed
at 1.25 $\mu$ in second order in combination with a blocking
filter.  The 200 mm camera was used with the SITe2 CCD and the slit
width was  $1''$, resulting in a spectral resolution of $\sim
1.2$ \AA\ (0.6 \AA/pixel).  The CCD was binned by two in the spatial
direction, giving a scale of $0.68''$/pixel.

The slit was rotated to coincide with the major axis of each galaxy. Target
galaxies were acquired by offsetting from nearby stars, but galaxies were
generally visible with the slit viewing optics.  Exposure times
were one hour, and a comparison lamp frame was taken at the same telescope
pointing immediately after each object exposure.  These comparison frames
were taken to track flexure of the spectrograph, but proved unnecessary.
For each frame, velocities are measured relative to
night sky lines of known wavelength on the frame.

In addition to major axis spectra, minor axis and intermediate slit positions
were obtained for a few galaxies in order to check for the presence of
non-circular motions (see Table 1).  Though only a few cases could
be tested, these were chosen for their suggestive morphologies. 
Apparently, LSB galaxies are fairly quiescent rotators
(see also O'Neil, Verheijen, \& McGaugh 2000).  Still, the frequency
and magnitude of noncircular motions in LSB galaxies merits further
investigation.

Immediately prior to the spectrographic portion of this observing run,
$R$-band images of target galaxies were obtained with the LCO 40 inch
telescope. These images were used to  determine inclinations and
position angles, and also provide an estimate of the exponential scale
length of the disk.  Unfortunately, a non-photometric sky
precluded using  these frames for surface photometry and mass modeling. 

\subsubsection{Kitt Peak}

Observations of galaxies from the sample of de Blok \etal\ (1996) and the
UGC were made with the KPNO 4 m in June 1999 and February 2000.
We used the RC-spectrograph and T2KB CCD with the KPC-24 grating (860
lines/mm) in second
order and the RG610 blocking filter.  The slit size was $1.5''$, yielding
$\sim 1.0$ \AA\ spectral resolution and a spatial scale of $0.69''$/pixel.
The available slit viewing equipment did not allow for confident
telescope positioning, so we measured offsets from nearby stars.
Repeated settings showed this to be a repeatable procedure.  We exercised
considerable care to insure that each galaxy was properly aligned on the
slit.  Cumulative exposure times were one hour. 

For galaxies with previous \HI\ observations, position angles were known
from the \HI\ reductions.  For the UGC galaxies, position angles are
measured from digital sky survey images. As with the Las Campanas spectra,
velocities are measured with respect to night sky lines on the same frame.
For the lowest surface brightness and lowest luminosity targets, we failed to
obtain high signal-to-noise emission line velocities from both Kitt Peak
and Las Campanas spectra.  Obtaining
good data for these interesting objects is possible, but
may be beyond the reach of 4 m class telescopes.  At both Kitt Peak
and Las Campanas, we were able to form rotation curves for
about 70\% of the galaxies with detectable emission.

\subsection{Rotation Curves}

Example of the two dimensional long slit spectra are shown in
Figure 1. The H$\alpha$ emission line, the strongest line, 
provides the highest signal-to-noise tracer of the
velocities.  [N II] $\lambda$6583\AA\ is also often measured.
The doublet [S II] $\lambda \lambda$6716,6732
is generally detected, often entangled in the
night sky OH lines beyond 6860 \AA.  The intensity of each [S II] line is
generally stronger than [N II] 6583 \AA, typical of low luminosity
galaxies (Rubin, Ford, \& Whitmore 1984). 
The strength of H$\alpha$ varies from object
to object, and thus also the accuracy of the rotation curve. 

\placefigure{Fig1}

Details of the velocity measuring procedure follow those given by
Rubin, Hunter, \& Ford (1991). Night sky OH lines are used for the 
two-dimensional wavelength calibration, so no rebinning is necessary.
For each spectrum, the velocity
zero-point is set by the night sky lines on that frame.
At successive  distances from the nucleus along the major axis,
velocities of both H$\alpha$ and [NII] (and occasionally [SII]) are
measured from the centroid of each emission line. 

Major axis line-of-sight velocities as a function of radius are shown 
in Fig. 2.  Mean velocities, formed from all measures within a small
radial bin, are shown along with their 1$\sigma$ errors determined from
the scatter in the measured velocities.  Where only one
point is measured within a bin, a nominal, conservative
1$\sigma$ error of 10 \kms\ is adopted.

\placefigure{Fig2}

The center of each galaxy is generally well defined by
the coincidence of the peak of the H$\alpha$ emission
with the peak of the stellar continuum emission from stars in the nucleus.  
The rotation curve is formed by flipping about this center and
superposing, by eye, the velocities from the
receding and approaching major axis and projecting to the plane of the
galaxy disk.  The resulting rotation curves are shown in Fig.\ 3 and
the data for those with Q = 1 are given in Table 2.

\placefigure{Fig3}

In forming the rotation curves for galaxies with fairly symmetrical
velocities, equal weight is given to the innermost and outermost
regions. Occasionally extinction  complicates the nuclear identification. 
Thus while the accuracy of a single
velocity point is about 4 \kms, we adopt an accuracy of 10 \kms\ for the
systemic velocity, due to the uncertainties in flipping the velocities to 
define \Vsys.  Mild asymmetries are quite common in
galaxies (Richter \& Sancisi 1994; Palunas \& Williams 2000).
These do not pose a problem to mass modeling (Paper II) provided proper
account is taken of the related uncertainties.  Gross asymmetries are
another matter:  such objects are excluded from further analysis.

\placetable{Table2}

A comparison of the central velocities derived here with systemic
velocities from various sources shows a good agreement. For 31 galaxies
with some independent recession velocity measurement,
the value of the mean (absolute) difference $\Delta V = 9.0 \kms$.
This is close to the nominal 10 \kms\ accuracy adopted for our values.
However, many of the external system velocities come from old catalog
values, so must contribute significantly to $\Delta V$.  

To make the most judicious use of the available data, we have constructed
hybrid rotation curves, which use \Ha\ data over the range of
radii where available, plus the
21 cm data to define the outermost points (Paper II). Such rotation curves
offer the resolution of the optical data and the extent of the 21 cm data.
The improved spatial resolution and smaller
error bars provide stronger constraints on mass models
than can be obtained from 21 cm data alone.

\section{Comparison of Optical and 21 cm Data}

A major goal of this study is to compare the high resolution optical rotation 
curves with curves derived from 21 cm studies.  We wish to investigate
whether the beam size of the \HI\ observations ($13''$ to $45''$)
degraded steeply rising rotation curves, causing them to
appear as slowly rising ones.  
The optical data points are overplotted on 21 cm
position-velocity diagrams (van der Hulst \etal\ 1993;
de Blok \etal\ 1996) in Figure 4.  In addition to the \Ha\ data presented
here, we also make use of the 
optical observations of 5 LSB galaxies by Swaters \etal\ (2000), of which
F568-3 is a duplicate. This results in 18 galaxies observed both at
\HI\ and H$\alpha$ for which a meaningful comparison can be made. This
represents a significant fraction of the two dozen galaxies in the
LSB \HI\ samples.

\placefigure{Fig4}

The agreement of the optical velocities with the \HI\ profiles varies on
a case by case basis.
The general shapes of many rotation curves are similar in \Ha\ and 21 cm,
but some are noticeably discrepant. In a few cases, the quality of the
optical curves are poor, so a meaningful comparison is not possible. 

Figure 5 shows the 21 cm rotation curves (van der Hulst \etal\ 1993;
de Blok \etal\ 1996) with the optical points superposed. 
The comparison sample has now decreased to 15, as 4 galaxies with poor
optical rotation curves are not included.
The 21 cm velocities generally extend to
larger radial distances than the optical.  The median value
of the ratio r(21cm)/r(\Ha) = 1.5.

\placefigure{Fig5}

Four of the galaxies we have observed have independent \Ha\ observations
from other sources.  F568-3 has been observed by us, and 
also by Pickering \etal\ (1998) and Swaters \etal\ (2000).  The
slow rise of its rotation curve, first indicated by 21 cm
observations (de Blok \etal\ 1996), has
been confirmed by all three independent optical data sets.
F563-1, F561-1, and UGC 5750 have been observed by  de Blok \&
Bosma (2002), with good general agreement with our observations
(Fig. 6).  In the case of F561-1, the error
bars are unusually large due to its low ($24^{\circ}$)
inclination combined with very limited emission.
Each of our points in this case is a single measure with
an adopted $\pm 10 \kms$
error, which projects to 25 \kms\ in the plane of the galaxy.  The
uncorrected data are in good agreement:  this is a repeatable
experiment.

\placefigure{Fig6}

The consistency of the \Ha\ and \HI\ data can be quantified by comparing
the measured velocities at small radii.
For 9 of the 15 galaxies (F561-1, F563-1, F568-1, F568-3, F571-V1,
F579-V1, F583-1, UGC 5750, and UGC 11557)
the velocity differences at 20$''$ are $\lesssim 8 \kms$.
Note that these are rotation velocities projected
to the plane of the galaxies; as observed on the plane of the sky,
they correspond to $V(r) \sin(i) < 5 \kms$.  Hence these
galaxies exhibit the same velocity rise that was observed at 21 cm. 

This can be seen directly in Figures 4 and 5.
In two cases (F561-1 and F571-V1) the agreement between the optical and
21 cm rotation curves is merely a matter of large error bars.  For the other
galaxies the agreement is genuine, and sometimes quite good (e.g., F568-3,
F583-1, and UGC 5750).  Indeed, the asymmetric structure of
F579-V1\footnote{De Blok \& McGaugh (1997) had already noted the difficulty
this case posed for attributing to beam smearing the slow rise of
the rotation curves derived from the 21 cm data.}
is apparent in both \Ha\ and 21 cm data, with particularly good
agreement between the two on the shape of the receding side (Fig. 4).

The remaining 6 galaxies
(F563-V2, F568-V1, F571-8, F574-1, F583-4, and UGC 6614)
have optical rotation curves which differ more noticeably
from the earlier \HI\ curves.  The reasons for this varies form case to case.
Unlike most of the LSB galaxies in this sample, UGC 6614 has a strong bulge
component for which one expects a rapidly rising then quickly falling
rotation curve.  This is apparent in the \Ha\ data, which hardly extend
into the range of disk-halo domination.  The 21 cm data do not probe
small radii, as there is a central hole in the HI distribution of this galaxy
(van der Hulst et al.\ 1993).
F571-8 is an edge-on galaxy, with the associated problems of optical
depth and projection effects (Matthews \& Wood 2001).
F574-1 and F583-4 do suffer from significant beam smearing as a result
of an elliptical beam shape which projects to a large physical size on
the galaxy.  This is convolved with the intrinsic HI distribution, which can
complicate matters further if emission is lacking from particular regions
(e.g., from a central hole).  The situation for F563-V2 and F568-V1 is less
clear as the optical data in these cases (from Swaters \etal\ 2000) has
substantial scatter.

The agreement or disagreement between the optical and 21 cm rotation
curves shows no correlation with LSB luminosity, apparent magnitude,
central surface brightness, or inclination. 
We do observe that the galaxy of lowest luminosity
(F565-V2; $M_B = -14.8$) has slowly rising velocities while the LSB
galaxy of highest luminosity (UGC 6614; $M_B = -20.3$) 
has steeply rising velocities.  This is consistent with the previously noted
relation between luminosity and rotation curve shape
(Rubin \etal\ 1985; Persic \& Salucci 1991; McGaugh \& de Blok 1998a).
Little further can be said about this here as 
the remaining galaxies are restricted to the fairly narrow
range $-18.8 < M_B < -16.5$.

Beam smearing effects are present in the 21 cm data, but are only
significant in a few cases.  Beam smearing has not caused us to mistake
steeply rising rotation curves for shallow ones.
There are rotation curves which do rise slowly,
and these are common in low luminosity and LSB galaxies.

\section{Scientific Impact}

With the new \Ha\ data for many objects which previously had been studied
in \HI, it is possible to assess the scientific impact of higher spatial
resolution.  Concerns have been expressed
that beam smearing in the \HI\ data might have seriously impacted the
derived shapes of the rotation curves (van den Bosch \etal\ 2000; Swaters
\etal\ 2000), and hence compromised conclusions which depend on these shapes.
There are two issues to which the initial rate of rise of the rotation
curve is particularly important:  maximal disks and cuspy halos.

One significant conclusion from the H$\alpha$ rotation curves is that 
one can now consider substantially higher maximum disk
stellar mass-to-light ratios (Swaters \etal\ 2000; Paper II)
than were inferred by de Blok \& McGaugh (1997).  A modest
change in the rate of rise of a rotation curve can lead to a large
change in the maximum disk \MLs.  For example, in the case of F583-1,
where only a small change in the shape of the rotation curve is found,
the maximum disk mass-to-light ratio rises from $\MLs^R = 1.5$ as determined
by the \HI\ curve (de Blok \& McGaugh 1997) to $\MLs^R = 6.5$ from the
\Ha\ data (Paper II).  This happens in spite of the rather modest change
in the input data, and this maximum disk mass-to-light ratio could be
pushed considerably higher still (to $\sim 12$) if one were to permit
consideration of a hollow halo or a modest overshoot of the innermost
data points (e.g., Palunas \& Williams 2000).  The maximum disk \MLs\
determined from the 21 cm data are already often uncomfortably high
compared to the expectations for stellar populations (de Blok \& McGaugh
1997).  As anticipated then, the H$\alpha$ data permit even more
implausible maximal values of \MLs.  Nevertheless, if one puts
little weight on stellar population mass-to-light ratios, it is
now formally possible to consider higher disk
mass-to-light ratios than were indicated by the 21 cm data alone.
The mass discrepancies of LSB galaxies are still large; this merely
transfers the missing mass from halo to disk.

The situation for cuspy (NFW) halos is less promising.
The \Ha\ data have a clear preference for dark matter halos
with constant density cores (Paper II; see also C\^ot\'e, Carignan \& Freeman
2000; Salucci 2001; Blais-Ouellette \etal\ 2001; de Blok \etal\ 2001)
rather than the cuspy cores produced in cosmological simulations with
cold dark matter (e.g., Navarro \etal\ 1997; Moore \etal\ 1999).
As we show in Paper II, this important scientific conclusion
is {\it not\/} an artifact of beam smearing.

\section{Conclusions}

We have presented H$\alpha$ rotation curves for a sample of LSB
galaxies.  This is a large increase in the number of LSB galaxies for
which such data are available.  The substantial mass discrepancies of
these galaxies make them useful probes of the dark matter problem
(de Blok \etal\ 2001; Paper II).

These data represent an order of magnitude improvement in
spatial resolution over 21 cm studies of the same galaxies (van der
Hulst \etal\ 1993; de Blok \etal\ 1996).  The optical data complement
the radio data.  The H$\alpha$ data define the inner rise of
the rotation velocities, and thus provide a more accurate determination of
the shape of the inner potential of the galaxy.  This is critical to
the question of whether dark matter halos have cusps or
cores. The 21 cm data extend to larger radii, probing the extent of
the dark matter halo and mapping out the gas, an important component
of the total baryonic mass in these systems. 

These new data allow us to address directly the concerns which have
been raised concerning the lower resolution of the radio data (van den 
Bosch \etal\ 2000; Swaters \etal\ 2000).  Beam smearing turns out
to be no problem for about half of the galaxies with both optical and 21
cm observations, and a serious problem in a only few.  
The basic scientific conclusions reached previously
(e.g., McGaugh \& de Blok 1998a,b) remain unaltered
as the systematic properties of the
rotation curves upon which these were based remain valid.
Indeed, they have become more clear in the improved data.

The most important result of these data are the tighter
constraints on mass models provided by the improved accuracy with
which the potentials have been traced.  This allows us to more clearly
distinguish between dark matter halo models with constant density
cores or central cusps in their density profiles.  These new 
high resolution data strongly disfavor the cuspy
halos predicted by cosmological simulations (Paper II).

\acknowledgements

We thank the observatories for providing telescope time, and their
staffs for the excellent level of support received.  We are grateful
to Rob Swaters and Renzo Sancisi for their comments, and to the
referee for a thorough examination of the data.
The work of SSM is supported in part by NSF grant AST9901663.


\clearpage
\input mcgaugh.tab1

\input mcgaugh.tab2

\clearpage
\begin{figure}
\figurenum{1}
\epsscale{0.9}
\plotone{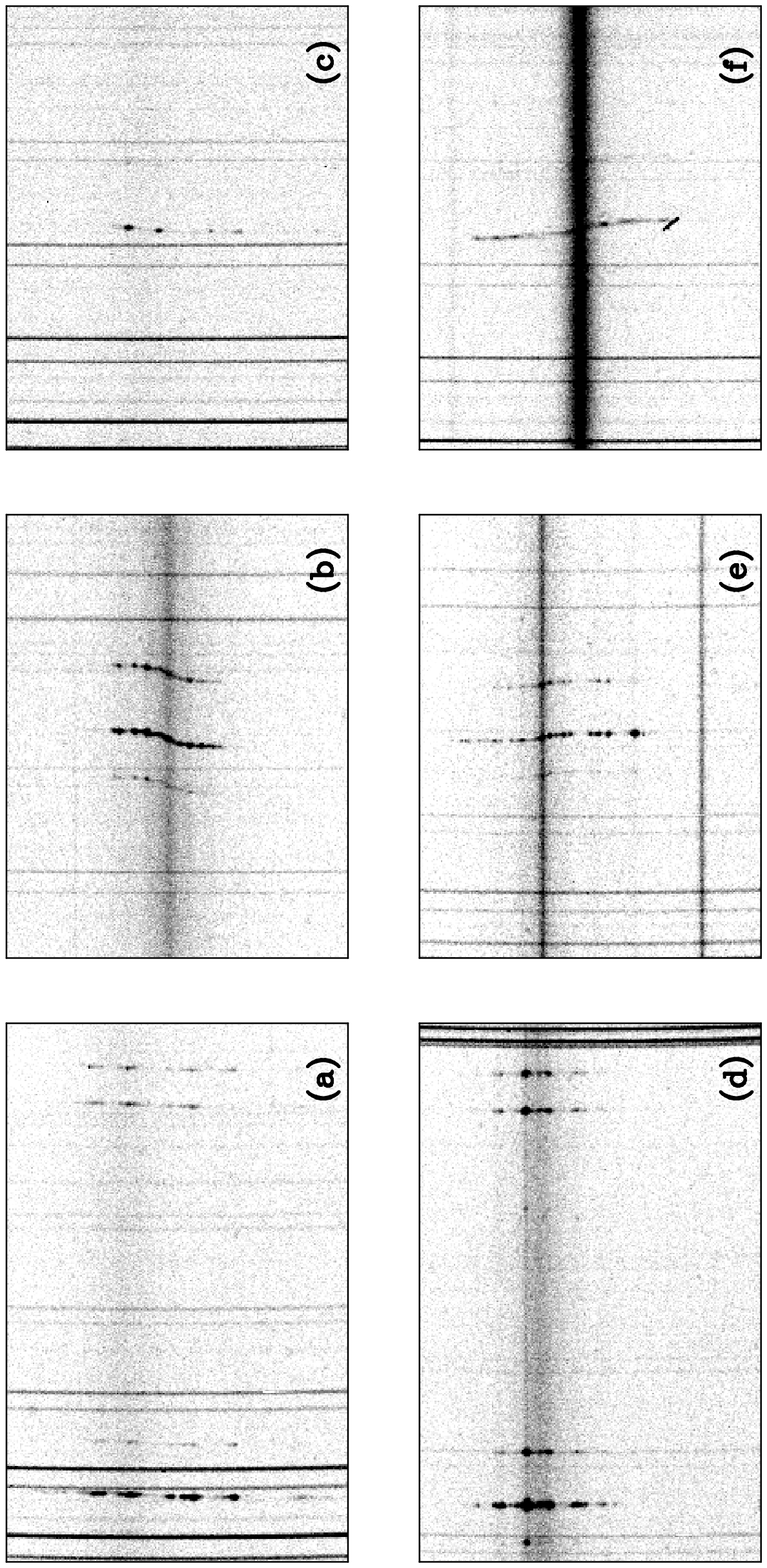}
\caption{The H$\alpha$ and [NII] spectral region in 6 galaxies. For
two, we show the spectrum including the [SII] lines.
a) ESO 1870510, Las Campanas 2.5 m telescope, 3600 s exposure. The weak [SII]
lines at $\lambda \lambda$ 6716, 6731 \AA\ are stronger than the [NII]
lines, characteristic of low luminosity galaxies (Rubin et al.\ 1984).
b) UGC 11616, 4 m telescope, 3600 s exposure.
The discreet knots reflect a knotty
morphology, and also very good seeing.
c) F563-1, 4 m telescope, 5400 s exposure, H$\alpha$ only.
Weak or absent [NII] is indicative of galaxies of low luminosity.
d) ESO 3520470,  Las Campanas 2.5 m telescope, 3600 s exposure. A Magellanic
irregular with relatively strong [SII]. The slit was placed along the bar,
but little or no rotation is observed.  No mass model was constructed
for this galaxy.
e) ESO 2060140, Las Campanas 2.5 m telescope, 3600 s exposure.
f) F571-8, 4 m telescope, 3600 s exposure.
A cosmic flaw decorates the bottom of H$\alpha$.}
\end{figure}

\clearpage
\begin{figure}
\figurenum{2}
\epsscale{0.9}
\plotone{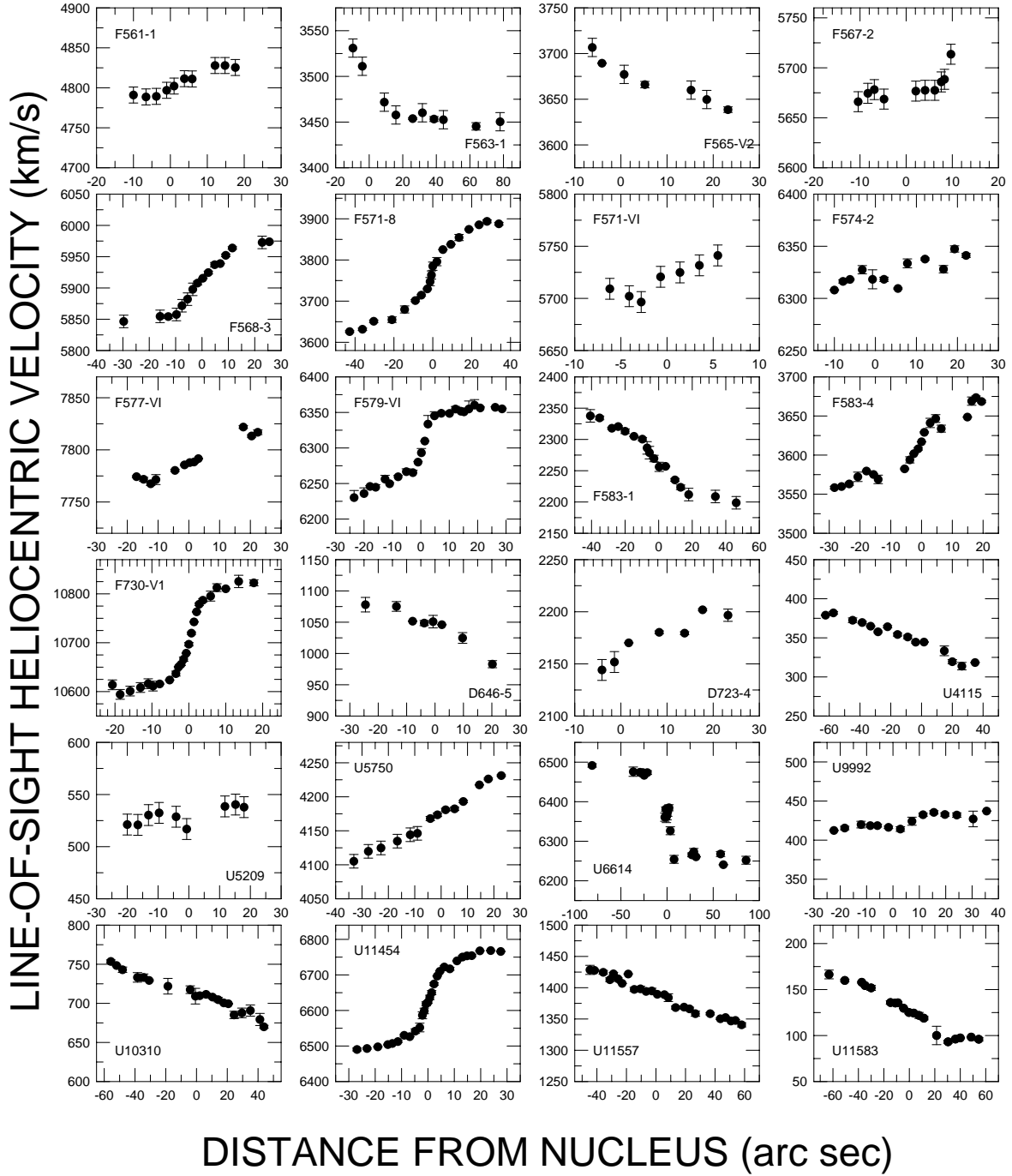}
\caption{Observed heliocentric
velocity plotted against angular distance from the center of each galaxy.
Each panel represents one galaxy, whose name is included.
F galaxies are from the LSB galaxy catalog
of Schombert \etal\ (1992), as studied in detail
by de Blok \etal\ (1996).  D galaxies are LSB dwarf
galaxies from the list of Schombert \etal\ (1997).
U galaxies are from the UGC (Nilson 1973).
Data for all of these were obtained with the KPNO 4 m.
E galaxies are from the ESO-LV (Lauberts \& Valentijn 1989) catalog
and were observed with the LCO 2.5 m.}
\end{figure}

\clearpage
\begin{figure}
\figurenum{2}
\epsscale{0.9}
\plotone{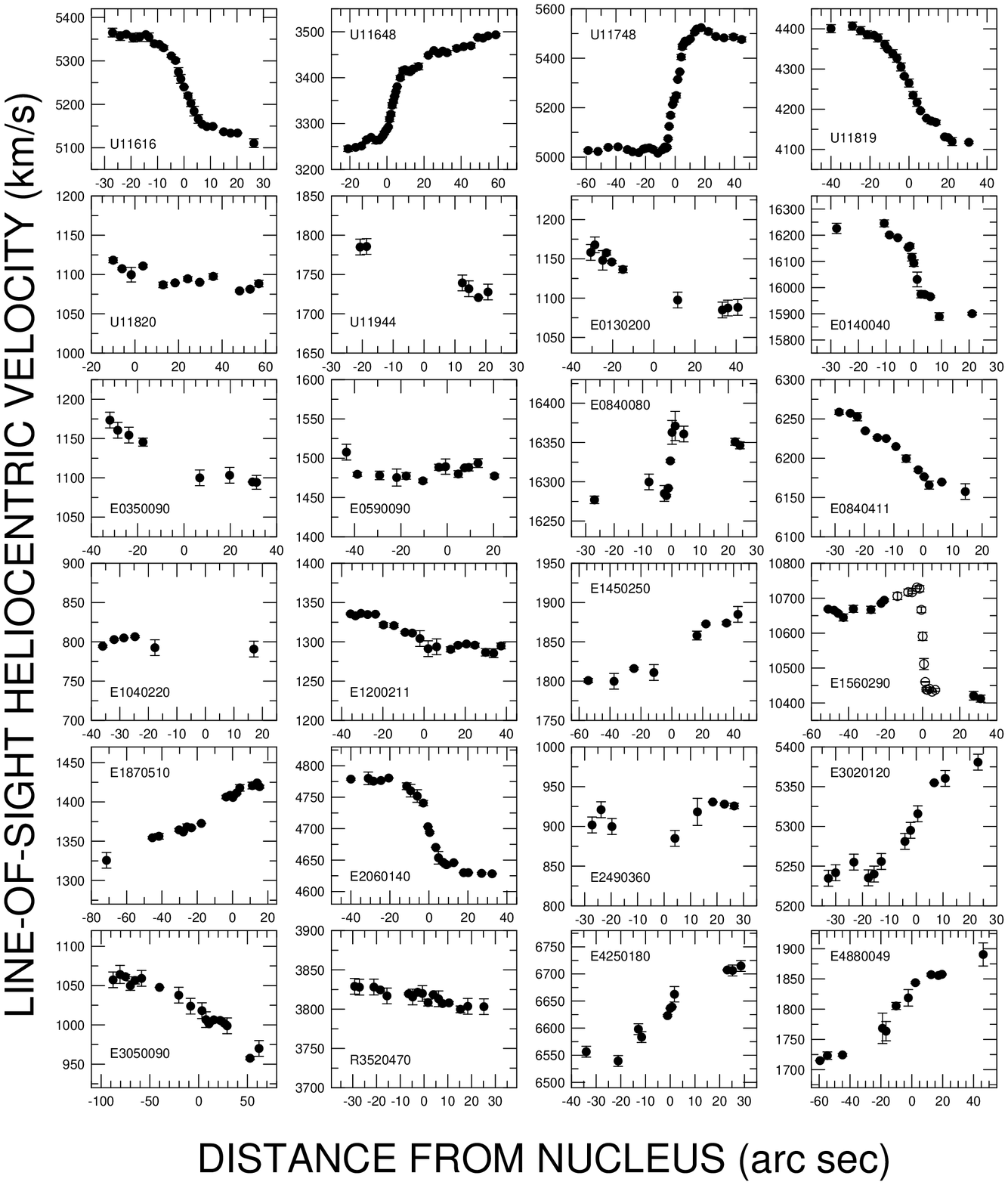}
\caption{continued.}
\end{figure}

\clearpage
\begin{figure}
\figurenum{3}
\epsscale{0.9}
\plotone{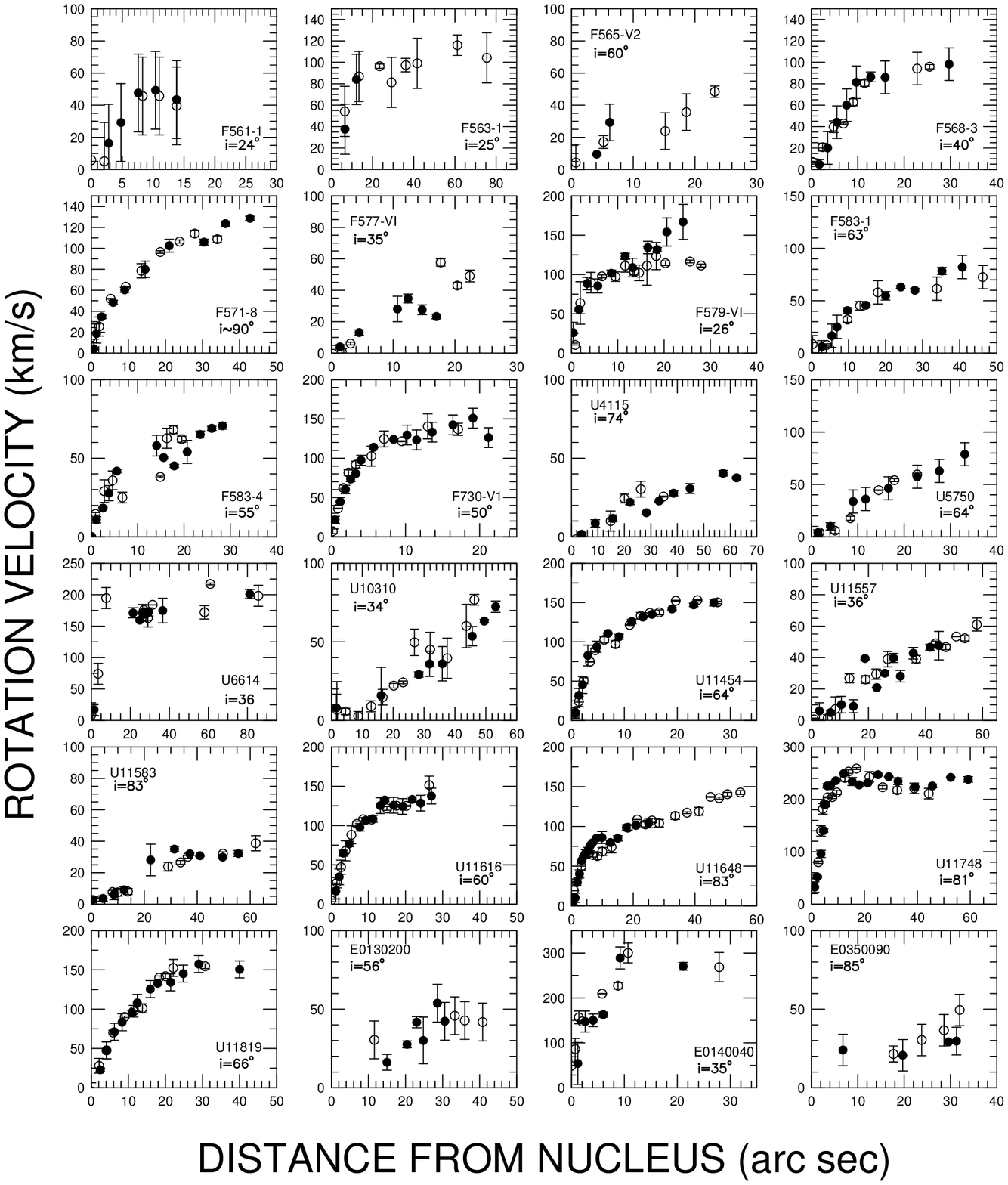}
\caption{The rotation curves
derived from the position-velocity diagrams of Fig.\ 2 by folding the data.
Rotational velocity, corrected for inclination and $(1+z)$,
is now centered on each systems' recession velocity.
Different sides of the galaxies are shown by different symbols.
Some galaxies have rotation curves which are well defined and
nicely symmetric (e.g., UGC 11819); others show mild asymmetry
(e.g., UGC 11748), while others remain ill-defined.
Note that in the case of ESO 1560290, no inclination correction has
been made and two different position angles are shown.}
\end{figure}

\clearpage
\begin{figure}
\figurenum{3}
\epsscale{0.9}
\plotone{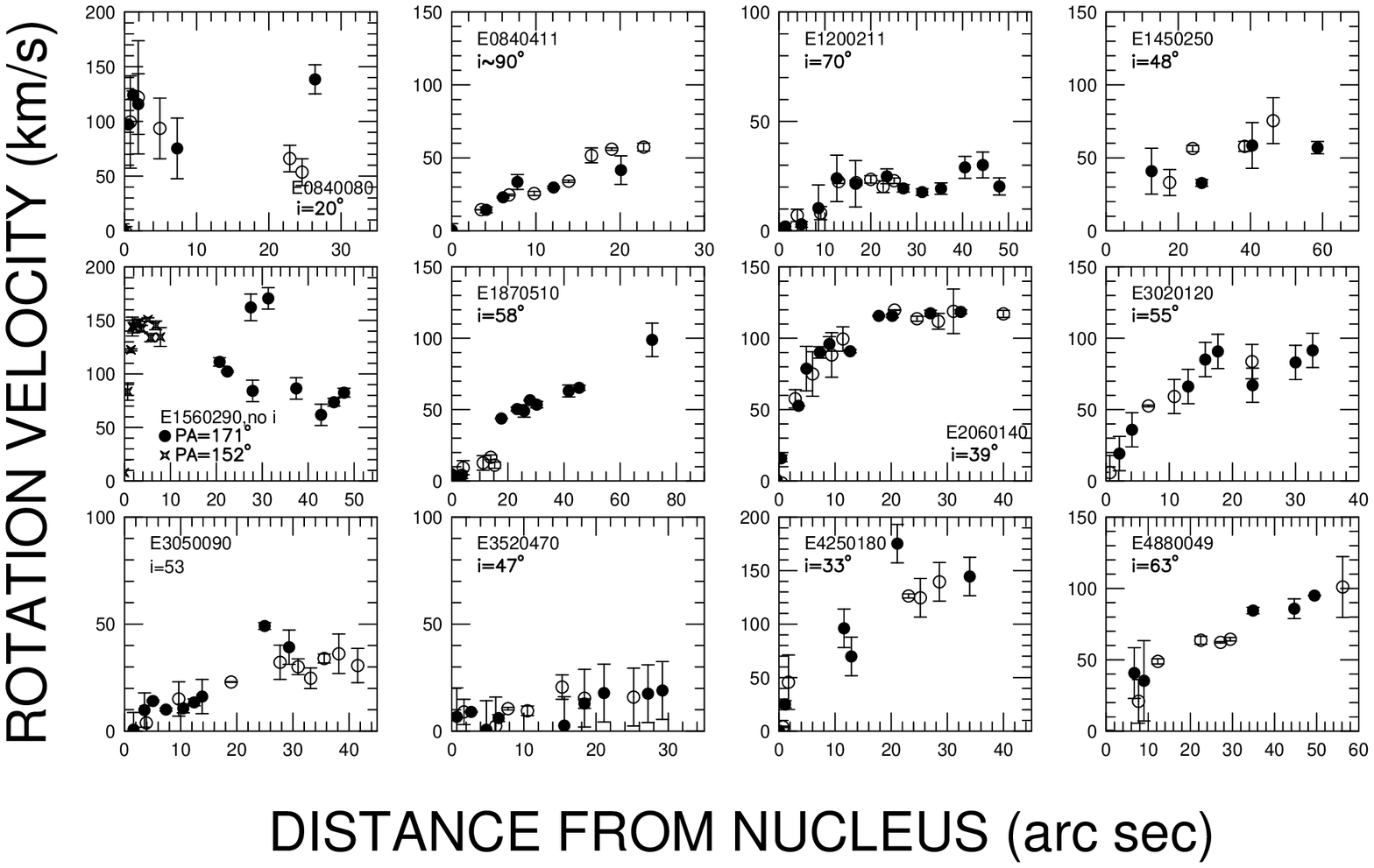}
\caption{continued.}
\end{figure}

\clearpage
\begin{figure}
\figurenum{4}
\epsscale{0.9}
\plotone{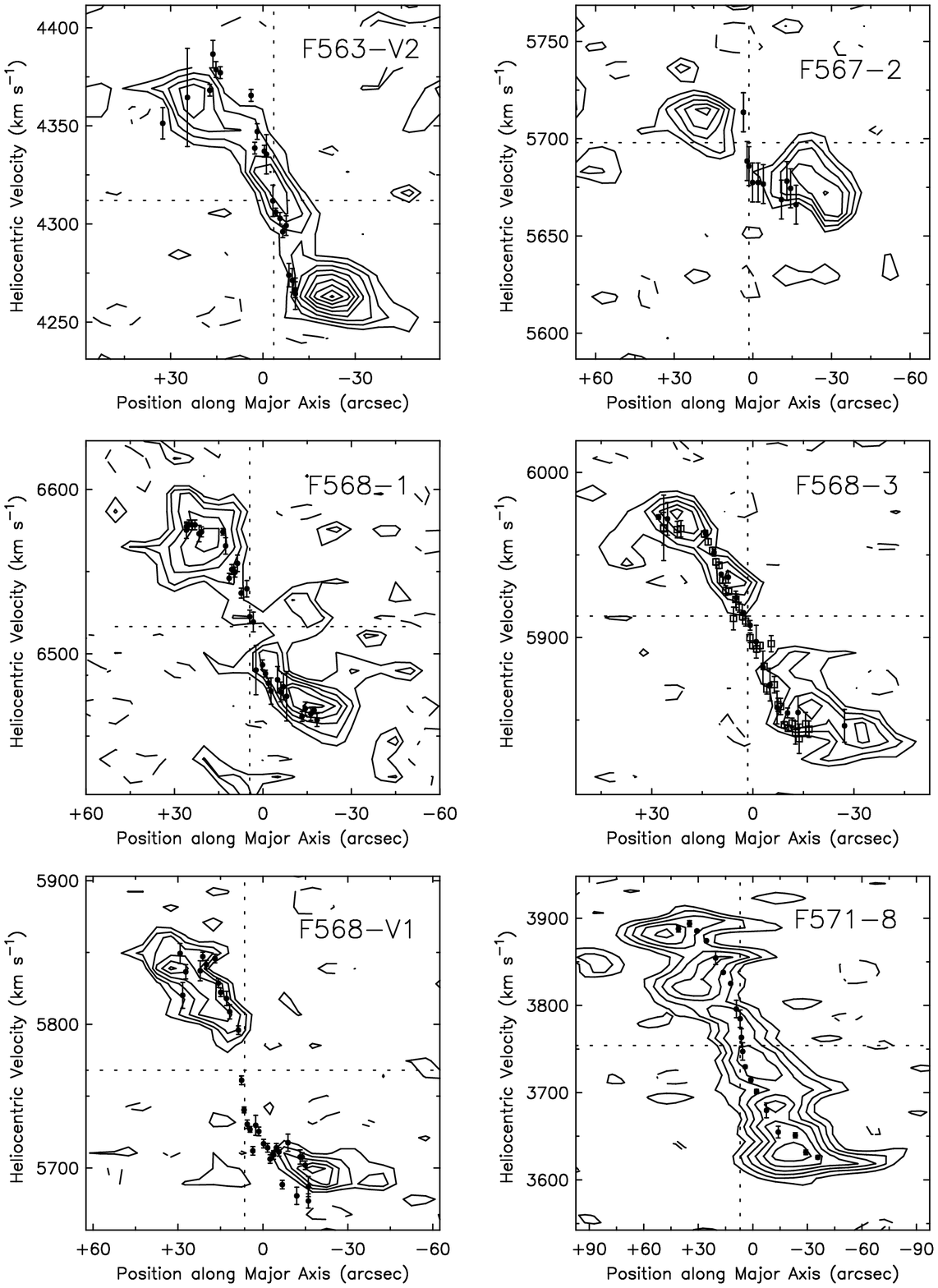}
\caption{Position-velocity diagrams
from the H$\alpha$ data (points with error bars)
overplotted on the 21 cm contours of
de Blok \etal\ (1996) and van der Hulst \etal\ (1993).
This allows direct comparison between the two data sets.}
\end{figure}

\clearpage
\begin{figure}
\figurenum{4}
\epsscale{0.9}
\plotone{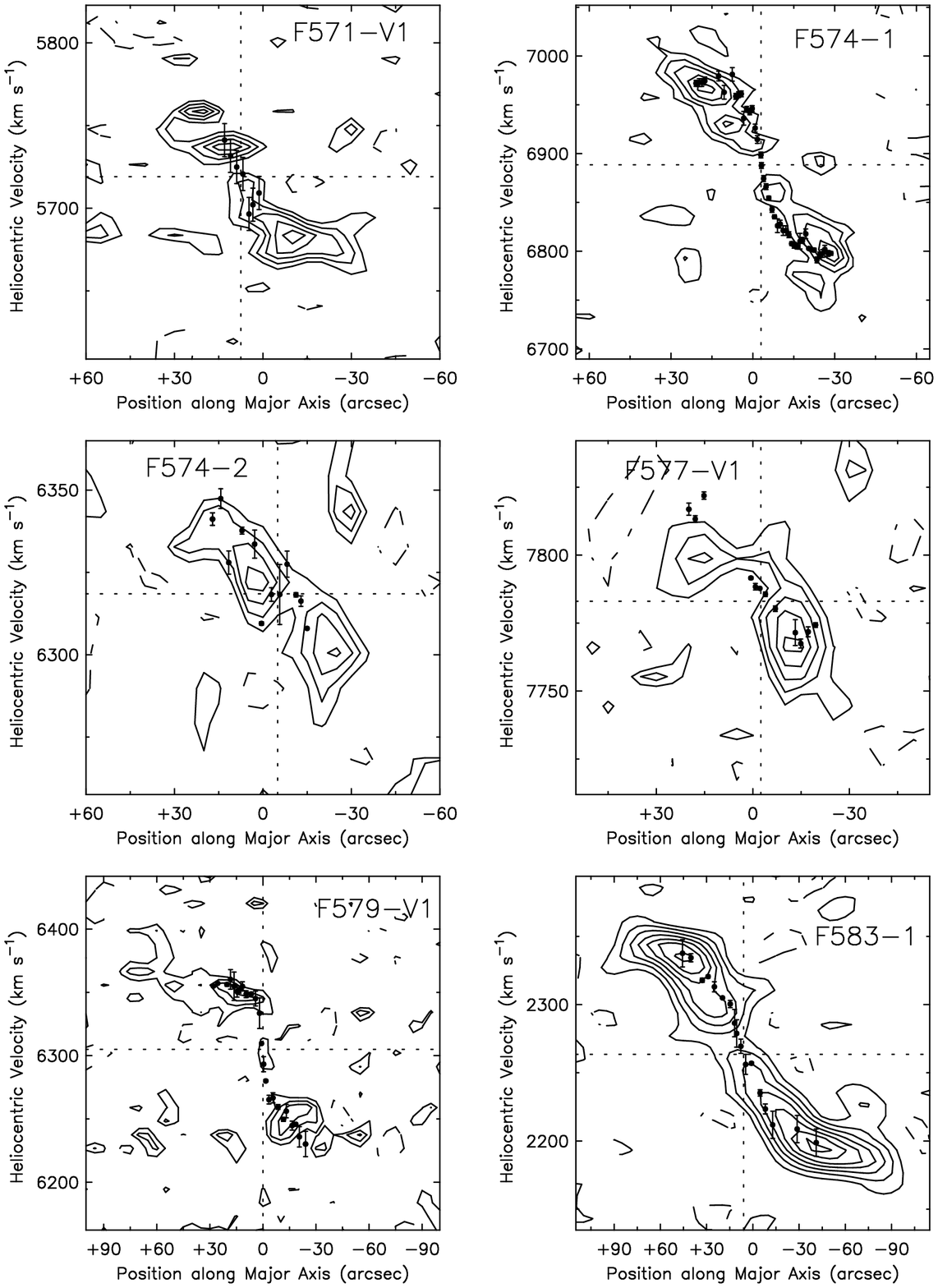}
\caption{continued.}
\end{figure}

\clearpage
\begin{figure}
\figurenum{4}
\epsscale{0.9}
\plotone{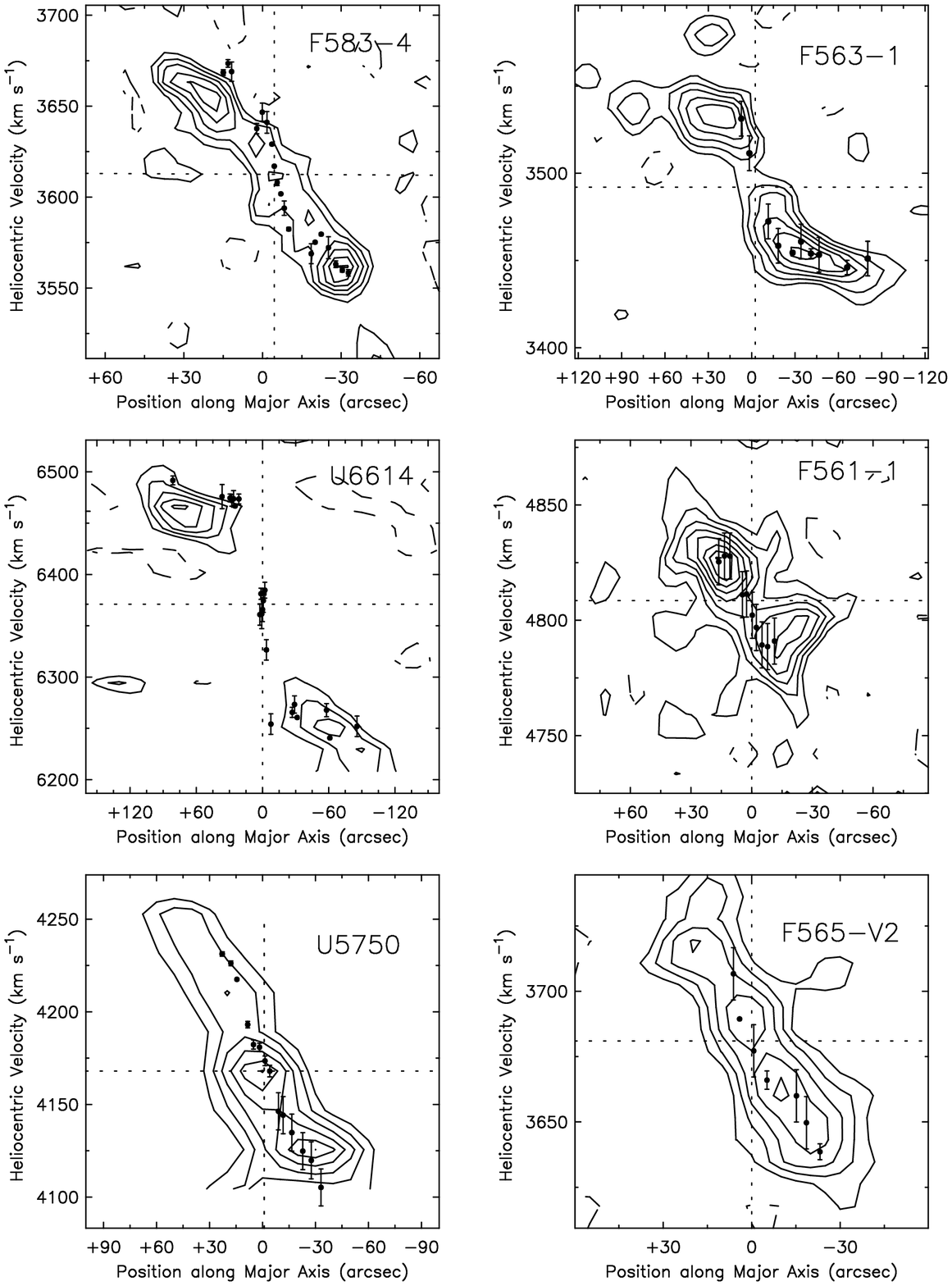}
\caption{continued.}
\end{figure}

\clearpage
\begin{figure}
\figurenum{5}
\epsscale{0.9}
\plotone{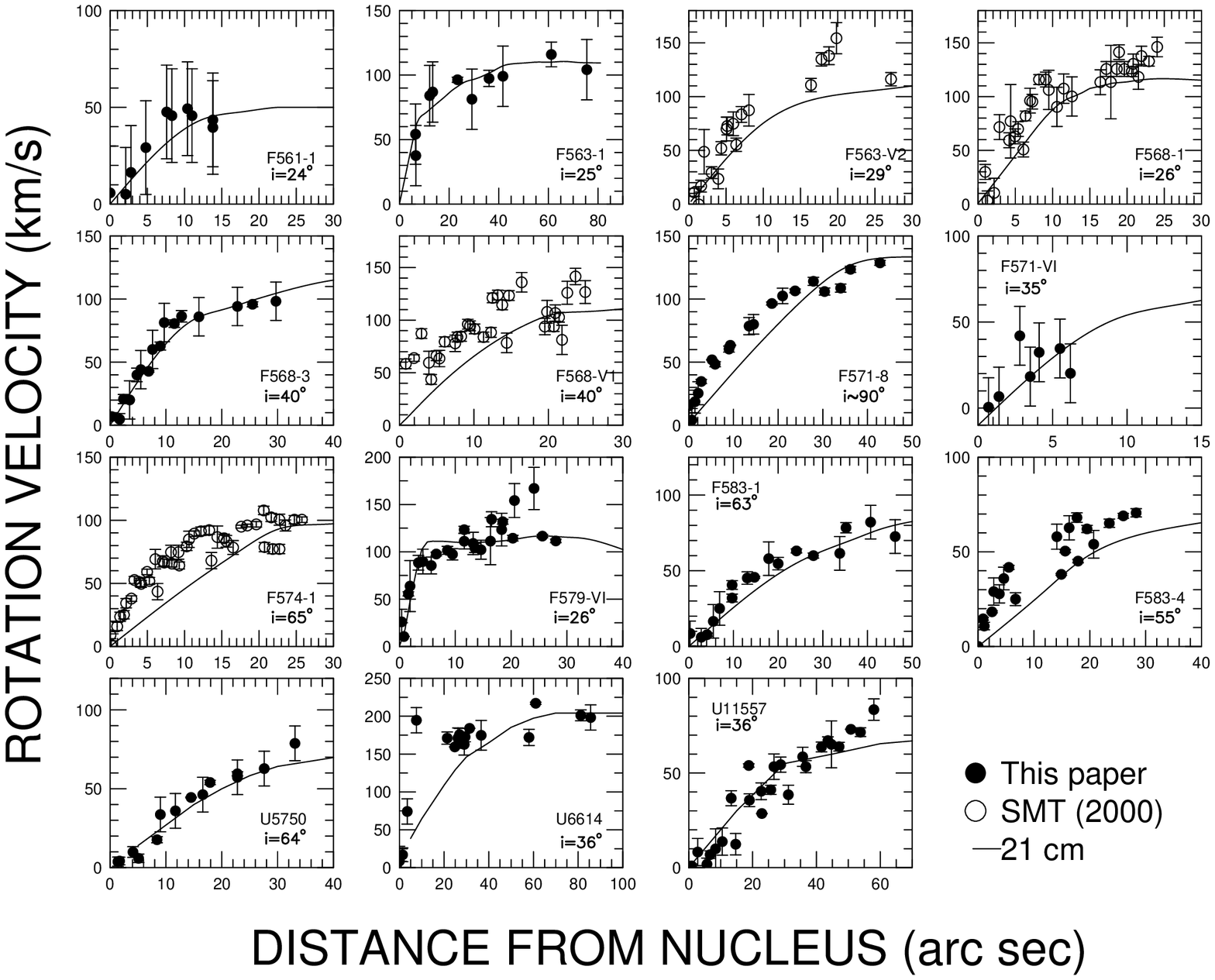}
\caption{The rotation
curves derived from the 21 cm
data of van der Hulst \etal\ (1993) and de Blok \etal (1996) are plotted
as solid lines together with the H$\alpha$ data (points with error bars).
In the case of UGC 11557 the 21 cm data are taken from Swaters (1999).
Different symbols denote the different sources of the \Ha\ data:
filled circles: this work; open circles: Swaters \etal\ (2000).
Cases of consistency and inconsistency are obvious from inspection,
and are discussed in the text.}
\end{figure}

\clearpage
\begin{figure}
\figurenum{6}
\epsscale{0.9}
\plotone{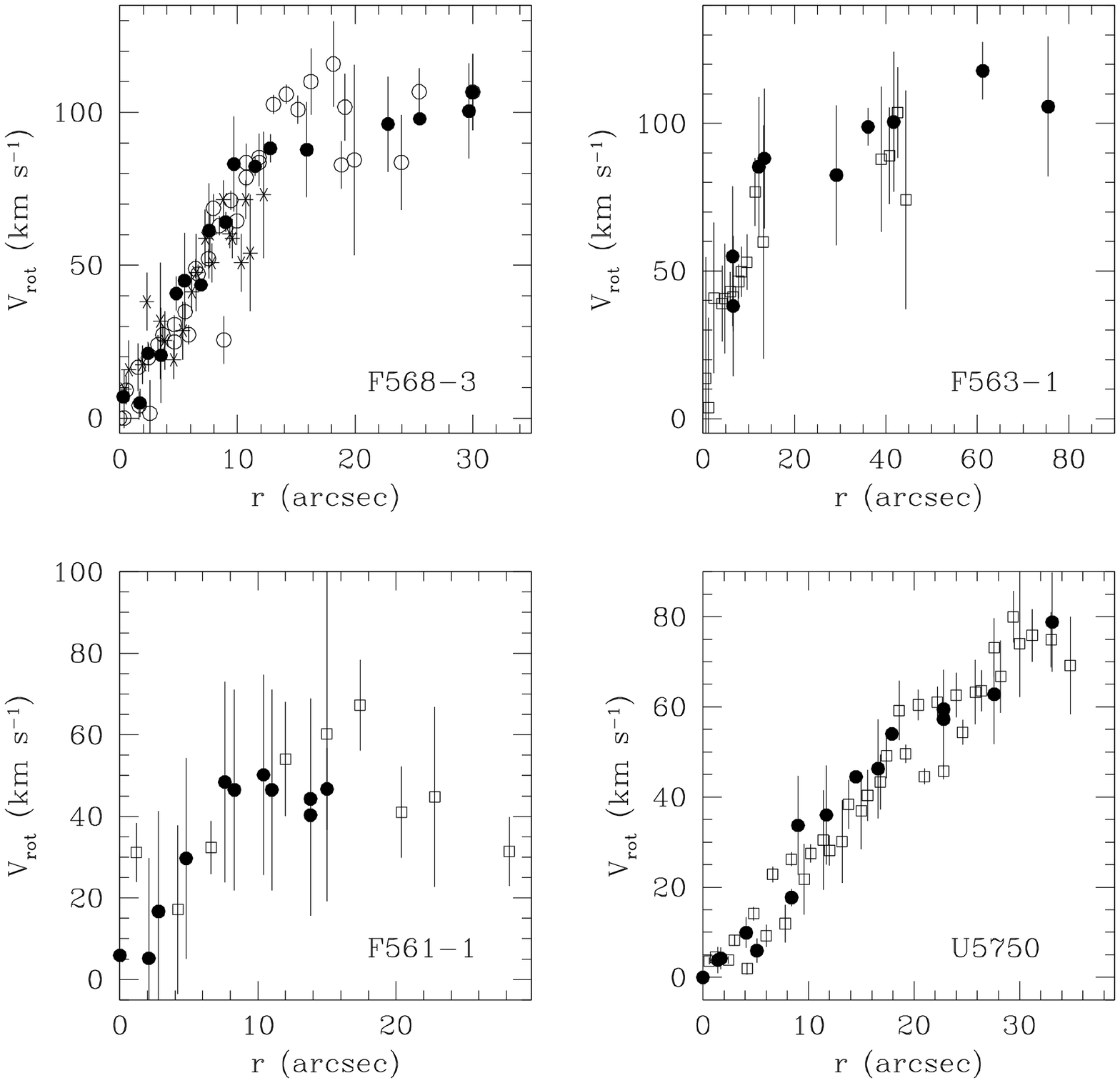}
\caption{The rotation curves derived from independent
H$\alpha$ observations for four LSB galaxies.  Solid circles are the data
presented here.  Open circles in
the top left panel (F568-3) are from Swaters et al.\ (2000) and
stars are from Pickering et al.\ (1998).  In the other panels,
open squares are from de Blok \& Bosma (2002).
Agreement between independent observations is good.}
\end{figure}

\end{document}

%% file: mcgaugh.tab1.tex
\begin{deluxetable}{lrrrrccl}
\renewcommand{\arraystretch}{.6}
\tablecaption{Observed Galaxies\label{Table1}}
\tablewidth{0pt}
\tablehead{
\colhead{Galaxy\tablenotemark{a}} & \colhead{$V_{\rm hel}$} &
\colhead{$i$} & \colhead{PA\tablenotemark{b}} & 
\colhead{Q\tablenotemark{c}} & \colhead{HI} & \colhead{Mass} &
\colhead{Comments} \\
\colhead{} & \colhead{(km s$^{-1}$)} & 
\colhead{$(^{\circ})$} & \colhead{$(^{\circ})$} &
\colhead{} & \colhead{} & \colhead{Model} & \colhead{}
}
\startdata
F561-1 	& 4809 & 24 & 55 & 2 & Y & N & bulge, faint disk \\ 
F563-1 	& 3502 & 25 & 161 & 1 & Y & Y & Mag.\ Irr\\ 
F563-V1\tablenotemark{d} & 3890 & 60 & 140 & 3 & Y & N & faint, bar\\
F565-V2	& 3681 & 60 & 20 & 2 & Y & N & ragged \\
F567-2	& 5675 & 20 & 119 & 3 & Y & N & faint \\
F568-3 	& 5913 & 40 & 169 & 1 & Y & Y & spiral with Mag.\ bar\\ 
F571-8 	& 3768 & $\sim$90 & 168 & 1 & Y & Y & edge-on\\ 
F571-V1	& 5721 & 35 & 45 & 2 & Y & N & faint, ragged\\ 
F574-2	& 6320 & 30 & 53 & 3 & Y & N & core, faint disk\\
F577-V1	& 7788 & 35 & 40 & 2 & Y & N & Mag.\ Irr\\
F579-V1	& 6305 & 26 & 120 & 1 & Y & Y & core, flocc. arms\\ 
F583-1 	& 2264 & 63 & 175 & 1 & Y & Y & Mag.\ Irr\\
F583-4 	& 3617 & 55 & 115 & 1 & Y & Y & fuzzy\\ 
F730-V1 & 10714 & 50 & 16 & 1 & N & Y & spiral\\
D646-5	& 1045 & \nodata & 29 & 2 & N & N & small; diffuse\\
D723-4	& 2175: & \nodata & 94 & 2 & N & N & small, irr\\
UGC 4115 & 343 & 74 & 140 & 1 & N & Y & = D631-7; knotty and diffuse \\
UGC 5209 & 530: & 53 & 158 & 3 & Y & N & fuzzy, small\\
UGC 5750 & 4177 & 64 & 167 & 1 & Y & Y & Mag.\ bar\\
UGC 6614 & 6371 & 36 & 108 & 1 & Y & Y & bulge; spiral ring\\
UGC 9992 & 430 & 30 & 35 & 2 & N & N & fuzzy, small\\
UGC 10310 & 724 & 34 & 19 & 2 & N & N & knotty spiral arms; slit missed nuc.?\\ 
UGC 11454 & 6628 & 64 & 106 & 1 & N & Y & fuzzy spiral, small core\\
UGC 11557 & 1390 & 36 & 94 & 1 & N\tablenotemark{e} & Y & fuzzy spiral, small core\\ 
UGC 11583 & 128 & 83 & 88 & 1 & N & Y & faint Mag.\ bar\\
UGC 11616 & 5244 & 60 & 99 & 1 & N & Y & fuzzy, Irr\\
UGC 11648 & 3350 & 83 & 145 & 1 & N & Y & Irr\\
UGC 11748 & 5265 & 81 & 103 & 1 & N & Y & Irr, bright core/bar?\\
UGC 11819 & 4261  & 66 & 167 & 1 & N & Y & fuzzy\\
UGC 11820 & 1100: & 50 & 48 & 2 & N & N & blobby, irr.\ streamers SW/NE, sp?\\
UGC 11944 & 1753 & 72 & 30 & 3 & N & N & bar\\
ESO 0130200 & 1123 & 56 & 124 & 2 & N & N & bulge, faint disk\\ 
ESO 0140040 & 16064 & 35 & 125 & 1 & N & Y & bulge, tight spiral arms\\ 
ESO 0350090 & 1124 & 85 & 24 & 3 & N & N & Irr\\ 
ESO 0590090 & 1486: & 30 & 132 & 3 & N & N & bulge, faint disk, knotty NE arm\\ 
ESO 0840080 & 16327 & 20 & 85 & 2 & N & N & bulge, faint disk, star supperposed\\ 
ESO 0840411 & 6200 & $\sim$90 & 11 & 1 & N & Y & edge-on\\ 
ESO 1040220 & 800: & 54 & 109 & 3 & N & N & blobby\\ 
ESO 1040440\tablenotemark{d} & 840: & 38 & 157 & 3 & N & N & irr, stars supperposed\\ 
ESO 1200211 & 1314 & 70 & 118 & 2 & N & Y & fuzzy Mag.\ bar\\ 
ESO 1450250 & 1837 & 53 & 128 & 2 & N & N & loose spiral\\ 
ESO 1560290 & 10583 & \nodata & 171,152 & 1 & N & N & spiral, bulge\\ 
ESO 1870510 & 1410 & 58 & 10 & 1 & N & Y & irr.\ spiral, floc.\\ 
ESO 2060140 & 4704 & 39 & 355 & 1 & N & Y & spiral\\ 
ESO 2490360 & 915: & 47 & 131 & 3 & N & N & Irr, blobby\\ 
ESO 3020120 & 5311 & 55 & 60, \phn 87 & 1 & N & Y & spiral, hint of bar?\\ 
ESO 3050090 & 1019 & 53 & 50, 140 & 1 & N & Y & barred spiral\\ 
ESO 3520470 & 3815 & 47 & 78 & 2 & N & N & Mag.\ Irr\\ 
ESO 4250180 & 6637 & 33 & 92 & 1 & N & Y & barred open spiral\\ 
ESO 4880049 & 1800 & 63 & 132 & 1 & N & Y & inclined Mag.\ bar\\ 
\enddata
\tablenotetext{a}{F, D, and U galaxies were
observed at Kitt Peak; E galaxies were observed at Las Campanas.}
\tablenotetext{b}{If multiple position angles are listed, the major axis
is given first.}
\tablenotetext{c}{H$\alpha$ rotation curve quality:
1 = good; 2 = fair; 3 = poor.}
\tablenotetext{d}{Low quality data not displayed in Fig.\ 2.}
\tablenotetext{e}{This galaxy was observed at 21 cm by Swaters (1999).}
\end{deluxetable}

%% file: mcgaugh.tab2.tex
\begin{deluxetable}{lrrr}
\renewcommand{\arraystretch}{.6}
\tablecaption{Rotation Curve Data\label{Table2}}
\tablewidth{0pt}
\tablehead{
\colhead{Galaxy} & \colhead{$R$} & \colhead{$V$} & \colhead{$\sigma_{V}$} \\
 \colhead{} & \colhead{(arcsec)} & \colhead{(km/s)} & \colhead{(km/s)}
}
\startdata
F583-1 & & & \\
&        -5.5   &     16.6    &    11.1 \\
&        -2.8   &      6.1    &     5.8 \\
&         0.3   &     -8.6    &     8.2 \\
&         4.1   &     -7.8    &     1.1 \\
&         9.7   &    -32.0    &     2.6 \\
\enddata
\tablecomments{The complete version of this table is in the electronic
edition of the Journal.  The printed edition contains only a sample.
These data are also available in electronic format from 
http://www.astro.umd.edu/$\sim$ssm/data \&
http://www.atnf.csiro.au/$\sim$edeblok/data.}
\end{deluxetable}

%% file: RCdataPP.bbl
\begin{references}
\reference{BBBdJHHLR} Bell, E.F., Barnaby, D., Bower, R.G.,
	de Jong, R.S., Harper, D.A., Hereld, M., Loewenstein, R.F., \&
                    Rauscher, B.J. 2000, \mnras, 312, 470
\reference{BdJTF} Bell, E.F., \& de Jong, R.S. 2001, \apj, in press
	(astro-ph/0011493)
\reference{BOAC} Blais-Ouellette, S., Amram, P., \& Carignan, C. 2001,
	\aj, in press (astro-ph/0006449)
\reference{BIMrev} Bothun, G., Impey, C., \& McGaugh, S. 1997, \pasp, 109, 745
\reference{CCF} C\^ot\'e, S., Carignan, C. \& Freeman, K.C. 2000, \aj, 120, 3027
\reference{CRix} Courteau, S., \& Rix, H.-W. 1999, \apj, 513, 561
\reference{DalcBern} Dalcanton, J.J., \& Bernstein, R.A. 2000, \aj, 120, 203
\reference{edBB} de Blok, W.J.G., \& Bosma, A., 2002, A\&A, submitted
\reference{edbhsblsb96} de Blok, W.J.G., \& McGaugh, S.S. 1996, \apjl, 469, L89
\reference{edbrot} de Blok, W.J.G., \& McGaugh, S.S., 1997, \mnras, 290, 533
\reference{edbmond} de Blok, W.J.G., \& McGaugh, S.S. 1998, \apj, 508, 132
\reference{edBssMaBvR} de Blok, W.J.G., McGaugh, S.S., Bosma, A., \&
	Rubin, V.C. 2001, \apjl, 552, L23
\reference{edBssMvR} de Blok, W.J.G., McGaugh, S.S., \& Rubin, V.C. 2001,
	in press (Paper II)
\reference{edbphot95} de Blok, W.J.G.,
        van der Hulst, J.M., Bothun, G.D., 1995, \mnras, 274, 235
\reference{edbbmh96}  de Blok, W.J.G., McGaugh, S.S., \&
	van der Hulst, J.M. 1996, \mnras, 283, 18
\reference{HSFRWH} Hoffman, G. L., Salpeter, E. E., Farhat, B., Roos, T.,
        Williams, H. \& Helou, G. 1996, \apjs, 105, 269
\reference{IBrev} Impey, C., \& Bothun, G. 1997, \araa, 35, 267
\reference{ISIBcat} Impey, C.D., Sprayberry, D., Irwin, M.J., \&
	Bothun, G.D. 1996, \apjs, 105, 209
\reference{K87} Kent, S. M. 1987, \aj, 93, 816
\reference{ESOLV} Lauberts, A., \& Valentijn, E.A. 1989,
	The surface photometry catalogue of the ESO-Uppsala galaxies,
	(Garching: European Southern Observatory) (ESO-LV)
\reference{MGphot} Matthews, L.D., \& Gallagher, J.S., III 1997,
	\aj, 114, 1899
\reference{MvD} Matthews, L. D., \& van Driel, W. 2000, A\&AS, 143, 421
\reference{MWood} Matthews, L. D., \& Wood, K. 2001, \apj, 548, 150
\reference{lsbdens} McGaugh, S.S. 1996, \mnras, 280, 337
\reference{mcgopt94} McGaugh S.S.,  Bothun G.D., 1994, \aj, 107, 530
\reference{mcgnodm98} McGaugh, S.S., \& de Blok, W.J.G., 1998a, \apj, 499, 41
\reference{mcgmond} McGaugh, S.S., \& de Blok, W.J.G. 1998b, \apj, 499, 66
\reference{MQGSL} Moore, B., Quinn, T., Governato, F., Stadel, J., \&
        Lake, G. 1999, \mnras, 310, 1147
\reference{NFW97} Navarro, J.F., Frenk, C.S., \& White, S.D.M.
	1997, \apj, 490, 493 (NFW)
\reference{ONBSredTF} O'Neil, K., Bothun, G.D., \& Schombert, J. 2000,
	\aj, 119, 136
\reference{ONBSCI} O'Neil, K., Bothun, G.D., Schombert, J., Cornell, M.E.,
	\& Impey, C.D. 1997, \aj, 114, 2448
\reference{ONVM} O'Neil, K., Verheijen, M.A.W., \& McGaugh, S.S. 2000,
	\aj, 119, 2154
\reference{UGCcat} Nilson, P. 1973, Uppsala general catalogue of galaxies,
	(Uppsala: Astronomiska Observatorium) (UGC)
\reference{PWFP} Palunas, P., \& Williams, T.B. 2000, \aj, 120, 2884
\reference{URC} Persic, M. \& Salucci, P. 1991, \apj, 368, 60
\reference{PCM} Pfenniger, D., Combes, F., \& Martinet, L. 1994, \aap, 285, 79
\reference{timrot97} Pickering, T.E., Impey, C.D., van Gorkom, J.H., \& 
	Bothun, G.D., 1997, \aj, 114, 1858
\reference{timUCSC} Pickering, T.E., Navarro, J.F., Rix, H.-W., \& Impey, C.D.
	1998, in Galactic Halos, ed. Zaritsky, D., ASP 136, 199
\reference{timU2936} Pickering, T.E., van Gorkom, J.H., Impey, C.D., \&
	Quillen, A.C., 1999, \aj, 118, 765
\reference{RSasym} Richter, O.-G.; Sancisi, R. 1994, \aap, 290, L9
\reference{RBFT} Rubin, V.C., Burstein, D., Ford, W.K., \& Thonnard, N. 1985,
        \apj, 289, 81
\reference{RFW} Rubin, V.C., Ford, W.K., Jr., \& Whitmore, B.C. 1984,
	\apjl, 281, L21
\reference{RHF} Rubin, V.C., Hunter, D.A., \& Ford, W.K., Jr. 1991,
	\apjs, 76, 153
\reference{paolo} Salucci, P. 2001, \mnras, 320, L1
\reference{LSBcat} Schombert, J.M., Bothun, G.D., Schneider, S.E., \&
        McGaugh, S.S. 1992, \aj, 103, 1107
\reference{Dgals} Schombert, J.M., Pildis, R.A., \& Eder, J.A. 1997,
	\apjs, 111, 233
\reference{SR} Sofue, Y. \& Rubin, V.C. 2001, \araa, in press
	(astro-ph/0010594)
\reference{SpSt} Spergel, D.N., \& Steinhardt, P.J. 2000, \prl, 84, 3760
\reference{spraytf95} Sprayberry D., Impey C.D., Bothun G.D., 
	\& Irwin M., 1995, \aj, 109, 558
\reference{Rob} Swaters, R.A. 1999, Ph.D. thesis, University of Groningen
\reference{SMT} Swaters, R.A., Madore, B.F., \&
	Trewhella, M., 2000, \apjl, 531, L107
\reference{vdBD} van den Bosch, F.C., \& Dalcanton, J.J. 2000, \apj, 534, 146
\reference{beamsmear} van den Bosch, F.C., Robertson, B.E.,
	Dalcanton, J.J., \& de Blok, W.J.G., 2000, \aj, 119, 1579
\reference{vdh93} van der Hulst J.M., Skillman E.D., Smith T.R.,
	Bothun G.D., McGaugh S.S.,  de Blok W.J.G., 1993, \aj, 106, 548
\reference{zwaantf95} Zwaan M.A., van der Hulst J.M., de Blok W.J.G.,
	\& McGaugh S.S., 1995, \mnras 273, L35
\end{references}
